\documentclass[12pt,aps,a4paper]{revtex4}
\usepackage{amsmath}
\usepackage{amssymb}
\usepackage{bm}
\usepackage{graphicx}
\begin{document}

\title{Calculations of the dynamical critical exponent
using the asymptotic series summation method}

\author{V.V.~Prudnikov\footnote{ E-mail\ :\ prudnikv@univer.omsk.su}, P.V.~Prudnikov and A.S.~Krinitsyn}

\affiliation{ Dept. of Theoretical Physics, Omsk State University 55a, Pr. Mira, 644077, Omsk, Russia}

\begin{abstract}
We consider how the Pad\'{e}--Borel, Pad\'{e}--Borel--Leroy, and conformal mapping summation methods for
asymptotic series can be used to calculate the dynamical critical exponent for homogeneous and disordered
Ising-like systems.\end{abstract}

\maketitle

\section{Introduction}

This paper shows in detail how the summation methods for asymptotic series can be used to calculate
the critical exponent $z$ determining the critical slowing down of the system relaxation time $\tau\sim\xi^{z}\sim|T-T_{c}|^{-z\nu}$ near
the temperature $T_c$ of the second-order phase transition $\xi$ is the correlation length and $\nu$ is the correlation
length exponent). The most complicated and interesting direction in the theory of critical phenomena is the
study of dynamical processes near a critical point. A significant achievement of the renormalization group
(RG) approach in studying the static critical phenomena is the constructed concept of universality classes
of the critical behavior of different systems characterized by similar critical properties. The universality
of the dynamical critical phenomena in contrast to the static phenomena is considerably weaker, which is
manifested in the existence of a wide range of models of the critical dynamics with different dynamical
critical behavior \cite{Hohenberg} and common critical properties in equilibrium. This difficulty is partially compensated
because the dynamical critical characteristics of many of these models can be expressed in terms of the
characteristics of their static critical behavior. This primarily concerns the models where the conservation
laws hold for the order parameter or any other long-lived excitations, in particular, the energy density \cite{Hohenberg}.
Because of the presence of preserved quantities, the purely hydrodynamic terms determining (and sometimes
suppressing) the influence of the fluctuation processes on the dynamical (relaxation) behavior can appear
in the stochastic equations of motion. The critical exponents describing the dynamical critical behavior can
be calculated in this case based on the static exponents with the same accuracy as the latter~\cite{Hohenberg}.

But there exist dynamical models such that the description of their critical behavior is not a trivial
problem. They are primarily models with purely dissipative equations of motion. To study the critical
behavior of such models, special methods must be developed, and dynamical quantities must be calculated
independently of the static quantities. Among such models, the Ginzburg--Landau dynamical model (model
$A$ in the Halperin--Hohenberg classification scheme \cite{Hohenberg}) is most interesting; this model was first proposed
by Landau and Khalatnikov to describe the anomalous sound attenuation in helium near a $\lambda$-point. The
dynamical critical behavior of other real systems such as the Ising-like magnets later came to be described
in the framework of this model.

The critical dynamics of model $A$ was studied in \cite{Bausch_DeDominicis_Halperin_1972_76} using the RG methods based on the $\varepsilon$-expansion.
To describe this dynamics, one of us first developed a field theory approach directly for the three- and twodimensional
homogeneous systems successively in the three-loop and four-loop approximations \cite{Prudnikov_1992}, \cite{Prudnikov_1997} and
then also for structurally disordered three-dimensional Ising-like systems in the three-loop approximation \cite{Prudnikov_1998}.
But the dynamical critical characteristics were determined in these papers with an accuracy considerably
less than the accuracy of the description of the static critical behavior of these systems in the six-loop
approximation of the theory in \cite{Baker,LeGuillou,Antonenko,Pelissetto}. This is primarily due to a fast increase in the volume of calculations
already for the lowest-order perturbation series, especially in the description of the critical dynamics of
structurally disordered systems. In this connection, the requirements for the accuracy of the summation
methods for the asymptotic series in this theory used to calculate the value of the dynamical critical
exponent $z$ in the description of the critical dynamics of model $A$ in homogeneous and disordered systems
become more rigid.
Here, we first present the results obtained by summing the series in this theory using the Pad\'{e}-Borel-
Leroy (PBL) and conformal mapping (CM) summation methods to calculate the values of the exponent
$z$. We compare the obtained results with experimental data and the results of a Monte Carlo computer
simulation of the critical dynamics.

\section{Model}

The field theory RG method, which permits calculating the values of the critical exponents characterizing
the asymptotic behavior of the thermodynamic and correlation functions near the critical temperature,
is widely used to describe the anomalous properties of the thermodynamic characteristics of systems in the
second-order phase transitions.
Our model of the critical behavior of a homogeneous ferromagnetic system is the classical spin system
thermodynamically equivalent to the $O(n)$-symmetric Ginzburg--Landau model with the effective Hamiltonian
\begin{equation} {\cal H}=\frac{1}{2}\int
d^{d}x\left(|\nabla\varphi|^{2}+r_{0}\varphi^{2}+\frac{g_{0}}{12}\varphi^{4}\right),
\label{f_H}
\end{equation}
where $d$ is the dimension of the system, $\varphi(\vec x,t)$ is the $n$-component order parameter (magnetization),
$r_{0}\sim{T-T_{c0}}$  ($T_{c0}$ is the critical temperature in the mean-field approximation), and $g_{0}>0$ is the vertex of
the magnetization fluctuation interaction. The dynamical behavior of a magnet in the relaxation regime
near a critical point in the framework of model $A$ is described by the Langevin-type kinetic equation for
the order parameter,
\begin{equation}
\frac{\partial\varphi}{\partial t}=-\lambda_{0}\frac{\delta\cal H}{\delta\varphi}+\zeta+\lambda_{0}h,
\label{f_dphi/dt}
\end{equation}
where $\lambda_{0}$ is the kinetic coefficient, $\zeta(\vec x,t)$ is a Gaussian random force, and $h(\vec x,t)$ is an external magnetic
field. It is well known that its solution in the form of correlation and response functions can be obtained
using a generating functional of the form
\begin{equation}
\Omega=\int D[\varphi]D[\tilde\varphi]\exp\left(-{\cal H}_{eff}[\varphi,\tilde\varphi]+\int(\varphi h+\tilde\varphi\tilde h)d^{d}xdt\right),
\label{f_Omega}
\end{equation}
where an auxiliary field $\tilde\varphi$ with the field source $\tilde h$ and the action functional
\begin{equation}
{\cal H}_{eff}=\int d^{d}xdt\left(\lambda_{0}^{-1}\tilde\varphi^{2}+i\tilde\varphi\left(\lambda_{0}^{-1}\frac{\partial\varphi}{\partial t}+\frac{\delta\cal H}{\delta\varphi}\right)\right).
\label{f_H_eff}
\end{equation}
are introduced. The functions of the order parameter response to the field $h$ are in this case determined as
\begin{equation}
G(x,t)=\left.\frac{\delta<\varphi(x,t)>}{\delta h(x,t)}\right|_{h=0}=\frac{1}{\Omega}\frac{\delta^{2}\Omega}{\delta h(x,t)\delta\tilde h(0,0)}=<\varphi(x,t)\tilde\varphi(0,0)>.
\label{f_G}
\end{equation}
Instead of the response function, it is more convenient to consider its vertex part $\Gamma^{(1,1)}(k,\omega)$.
The long-range long-lived fluctuations in the order parameter, which mainly determine the anomalies
of the equilibrium and nonequilibrium characteristics of systems as $T\to T_{c}$, are taken into account in
the framework of the RG methods developed in \cite{Hohenberg,Zinn-Justin}.
The $(N+\tilde N)$th-order renormalized dynamical
vertex functions $\Gamma_{R}^{(N,\tilde N)}(k,\omega)$, which are used in the theory and uniquely determine all the observable
characteristics of the system, are given by the RG differential equation
\begin{equation}
\left[\mu\frac{\partial}{\partial\mu}+\beta\frac{\partial}{\partial g}-r\gamma_{r}\frac{\partial}{\partial r}-\lambda\gamma_{\lambda}\frac{\partial}{\partial\lambda}-\frac{(N+\tilde N)}{2}\gamma_{\varphi}\right]\Gamma_{R}^{(N,\tilde N)}=0
\label{f_RGU_g}
\end{equation}
with the renormalized charge $g$, the reduced temperature $r$, and the kinetic coefficient $\lambda$. The asymptotic
behavior of the thermodynamic and correlation functions, which have a power-law character as $T\to T_{c}$,
is determined by the zero $g^{*}$ of the function $\beta(g)$: $\beta(g^{*})=0$, and by the function values $\gamma_{r}(g^{*})$, $\gamma_{\lambda}(g^{*})$ and
$\gamma_{\phi}(g^{*})$ specifying the static and dynamical critical exponents:
$\nu=(2+\gamma_{r}(g^{*}))^{-1}$, $\eta=\gamma_{\varphi}(g^{*})$, $z=2+\gamma_{\lambda}(g^{*})$
(for example, if $\gamma=\nu(2-\eta)$, then the magnetic susceptibility behaves as $\chi\sim|T - T_{c}|^{-\gamma}$).

The functions $\beta(g)$, $\gamma_{r}(g)$, $\gamma_{\lambda}(g)$ and $\gamma_{\phi}(g)$ contained in the differential RG equation can be calculated
as series in $g$. If the space dimension $d$ is close to four, then the coordinate of the fixed point $g^{*}$ of
the function $\beta(g)$ takes small values. The methods of the perturbation theory in the coupling constant
$g\sim 4-d$ can be used in this case, and the critical exponents can be calculated. For real systems with
$d=3,2$ the series in $g$ are asymptotic series, and they can be summed using special methods not based on
the perturbation theory representations \cite{Baker,LeGuillou}.

In the description of the critical behavior of structurally disordered systems with frozen nonmagnetic
admixture atoms or vacancies at the lattice nodes, which play the role of point defects of the structure, the
additional term
\begin{equation}
\Delta{\cal H}[\varphi,V]=\frac{1}{2}\int d^{d}xV(x)\varphi^{2},
\label{f_DH}
\end {equation}
where $V(x)$ is the potential of the random field of defects, is introduced in the effective Hamiltonian of
Ginzburg--Landau model (\ref{f_H}), which leads to fluctuations in the local critical temperature $r_{0}\sim{T-T_{c0}}$.
The distribution of the structure defects over the volume of the system is assumed to be Gaussian (taking
the deviations from the Gaussian distribution into account gives only corrections inessential in the critical
domain) with the distribution function
\begin{equation}
P[V]=A_{V}\exp\left[-(16v_{0})^{-1}\int d^{d}xV^{2}(x)\right],
\label{f_P}
\end{equation}
where $A_{V}$ is the normalization factor and $v_{0}$ is a positive parameter proportional to the concentration of
defects and to their squared potential. The correlation functions and the response functions for structurally
disordered systems, obtained later using the generating functional, must be averaged over the random defect
potential $V(x)$. This averaging can be performed most effectively using the replica method, which is widely
used for similar purposes in the study of weakly disordered systems (see, e.g., \cite{Dotsenko}). The essence of this
method is that the following mathematical transformation is formally used to obtain the free energy of a
disordered system in averaging over the defect distribution:
\begin{equation}
-F/T=\left<\!\left<\ln Z\right>\!\right>=\lim_{m\to 0}\frac{1}{m}\ln\left<\!\left<Z^{m}\right>\!\right>=\lim_{m\to 0}\frac{1}{m}\ln{\rm Sp}_{\{\varphi_{i}\}}\exp(-\hat H_{\rm repl}[\{\varphi_{i}\}]/T),
\label{f_F/T}
\end{equation}
where the double angle brackets denote the average over the probability distribution $P[V]$ of different defect
configurations, $Z$ is the partition function of the original disordered system, and $m$ replicas ("images") of
the original field  $\varphi$ - $\{\varphi_{i}\}$ are introduced with the replica indices $i=1,\ldots,m$. The replica Hamiltonian is
then determined as
\begin{equation}
\exp(-\hat H_{\rm repl}^{(m)}[\{\varphi_{i}\}]/T)=\left<\!\!\left<\prod\limits_{i=1}^{m}\exp(-\hat H[\{\varphi_{i}\}]/T)\right>\!\!\right>,
\label{f_exp}
\end{equation}
and is translation invariant, in contrast to the original Hamiltonian. Applying this averaging procedure to
dynamical generating functional (\ref{f_Omega}), we can obtain the replica action functional
\begin{eqnarray}
\hat H_{\rm repl}^{(m)}&=&\sum_{i}\int d^{d}xdt\left[-\lambda_{0}^{-1}\tilde\varphi_{i}\tilde\varphi_{i}+i\tilde\varphi_{i}\left(\lambda_{0}^{-1}\frac{\partial\varphi_{i}}{\partial t}-\nabla^{2}\varphi_{i}+r_{0}\varphi_{i}\right)+\frac{i}{3!}g_{0}\tilde\varphi_{i}\varphi_{i}\varphi_{i}\varphi_{i}\right]+\nonumber\\
&+&4v_{0}\sum_{i,j}\int d^{d}xdtdt'\tilde\varphi_{i}(x,t)\varphi_{i}(x,t)\tilde\varphi_{j}(x,t')\varphi_{j}(x,t').
\label{f_H_repl}
\end{eqnarray}
instead of the action functional for homogeneous model (\ref{f_H_eff}).
According to the replica method, the properties
of the original disordered system are obtained in the limit as $m\rightarrow 0$.
This limit annihilates the connected
diagrams containing loops formed by the admixture vertex $v_{0}$.

Taking the structural disorder of systems into account in describing the critical
behavior thus leads to introducing an additional interaction vertex $v_{0}$ in the
effective Hamiltonian. This vertex determines the effects of interaction of fluctuations
of the $nm$-component order parameter in terms of the field of defects.
The subsequent procedure for renormalizing the dynamical vertex functions is given by
the RG differential equation
\begin{equation}
\left[\mu\frac{\partial}{\partial\mu}+\beta_{g}\frac{\partial}{\partial g}+\beta_{v}\frac{\partial}{\partial v}-r\gamma_{r}\frac{\partial}{\partial r}-\lambda\gamma_{\lambda}\frac{\partial}{\partial\lambda}-\frac{(N+\tilde N)}{2}\gamma_{\varphi}\right]\Gamma_{R}^{(N,\tilde N)}=0
\label{f_RGU_gv}
\end{equation}
with the renormalization charges $g$ and $v$.
The functions $\beta_{g}$ and $\beta_{v}$ in this case depend on $g$ and $v$ as on
model parameters.
The fixed points $(g^{*},v^{*})$ of the RG transformations are determined by the zeros of
the functions $\beta_{g}$ and $\beta_{v}$, $\beta_{g}(g^{*},v^{*})=0$, $\beta_{v}(g^{*},v^{*})=0$,
and the critical exponents are determined by the values of the functions
$\gamma_{r}(g^{*},v^{*})$, $\gamma_{\varphi}(g^{*},v^{*})$ and $\gamma_{\lambda}(g^{*},v^{*})$
at the corresponding stable fixed points of the RG transformation.

As in the case of homogeneous systems, the functions $\beta_{g}$, $\beta_{v}$, $\gamma_{r}$, $\gamma_{\lambda}(g)$ and $\gamma_{\phi}(g)$
can be calculated as series in $g$ and $v$ only if the system dimension $d$ is close to four.
For real systems with $d = 3$, the series in $g$ and $v$ are only asymptotically convergent.

\section{Summation methods for asymptotic series}

To discuss the dynamical critical behavior of homogeneous systems and
to calculate the exponent $z$, we only need the functions $\beta(g)$ and $\gamma_{\lambda}(g)$.
The explicit form of the first of them in the six-loop approximation
for three-dimensional systems and in the four-loop approximation for two-dimensional systems
was obtained in \cite{Baker,LeGuillou}.
We calculated the dynamical scaling functions $\gamma_{\lambda}$ in the four-loop approximation
for two- and three-dimensional Ising models \cite{Prudnikov_1997}.
As a result, the functions $\beta(g)$ and $\gamma_{\lambda}(g)$ can be represented as the
series in the charge $g$
\begin{equation}
\frac{\beta(g)}{g}=-1+g-0.716174g^{2}+0.930767g^{3}-1.582388g^{4},
\label{f_gamma_d=2}
\end{equation}
$$
\gamma_{\lambda}(g)=0.027053g^{2}-0.004184g^{3}+0.022130g^{4},
$$
for the two-dimensional model and
\begin{equation}
\frac{\beta(g)}{g}=-1+g-0.422497g^{2}+0.351070g^{3}-0.376527g^{4}+0.495548g^{5}-0.749689g^{6},
\label{f_gamma_d=3}
\end{equation}
$$
\gamma_{\lambda}(g)=0.008399g^{2}-0.000045g^{3}+0.020423g^{4},
$$
for the three-dimensional model.
Based on the method proposed by Lipatov \cite{Lipatov}, it was shown in \cite{Bresin} that
although the general term of the series for the function $\beta(g)$ increases factorially,
\begin{equation}
\beta(g)=\sum\limits_{n=0}^{\infty}c_{n}g^{n},\qquad c_{n}\approx c(-a)^{n}n^{b}n![1+O(1/n)],
\label{f_beta}
\end{equation}
the series nevertheless satisfies the asymptotic convergence condition
\begin{equation}
\left|\beta(g)-\sum\limits_{n=0}^{N}c_{n}g^{n}\right|\le C\sigma^{N+1}[(N+1)!]^{\sigma}|g|^{N+1}
\label{f_ASU}
\end{equation}
where $C$ and $\sigma$ are some constants, in the wedge-shaped domain $|arg\ g|\le\frac{\pi}{2}\sigma$
on the complex plane. Matching asymptotic expansion (\ref{f_beta}) and the values of the first-order
coefficients gives information about all the terms of the series and permits reconstructing
the function $\beta(g)$ approximately, but this requires
a special summation procedure for asymptotically convergent series. Special summation methods
for such series were developed in \cite{Baker,LeGuillou,Antonenko,Pelissetto,Hardi,Kazakov,Suslov};
the most effective of them are the Pad\'{e}-Borel (PB), PBL, and CM methods.

In \cite{Honkonen}, the Lipatov technique for estimating higher-order terms in this theory
was generalized to the critical dynamics, and it was proved that the Lipatov asymptotic
expansion also holds for the dynamical RG function $\gamma_{\lambda}$.
Therefore, the above summation methods, which were successfully used to analyze the
static RG functions, can also be used to analyze the dynamical RG functions.

For the three-dimensional disordered Ising model, the obtained representations
of the functions $\beta_{g}$, $\beta_{v}$ and $\gamma_{\lambda}$ as series in the
charges $g$ and $v$ have the forms
$$
\frac{\beta_{g}}{g}=-1+g+\frac{3}{2}v-\frac{308}{729}g^{2}-\frac{104}{81}gv-\frac{185}{216}v^{2}+\sum\limits_{i+j\geq 3}b_{i,j}^{(g)}g^{i}v^{j},
$$
\begin{equation}
\frac{\beta_{v}}{v}=-1+\frac{2}{3}g+v-\frac{92}{729}g^{2}-\frac{50}{81}gv-\frac{95}{216}v^{2}+\sum\limits_{i+j\geq 3}b_{i,j}^{(v)}g^{i}v^{j},
\label{f_gamma}
\end{equation}
$$
\gamma_{\lambda}=\sum\limits_{i,j=0}^{3}\gamma_{i,j}g^{i}v^{j}=-\frac{1}{4}v+0.008400g^{2}+0.030862gv+0.053240v^{2}-\\
$$
$$
-0.012642g^{3}-0.041167g^{2}v-0.152964gv^{2}-0.049995v^{3},
$$
where the coefficients $b_{i,j}^{(g)}$, $b_{i,j}^{(v)}$ and $\gamma_{i,j}$ were calculated
in \cite{Pelissetto} and \cite{Prudnikov_1998} in the respective three-loop and
six-loop approximations.

As a preliminary illustration of the PB, PBL, and CM summation methods, we consider the
exactly solvable problem of determining the ground state energy of the linear anharmonic
oscillator with the Hamiltonian
\begin{equation}
{\cal H}=p^{2}+x^{2}+gx^{4}.
\label{f_H_osts}
\end{equation}
This permits detecting special features of the use of each of the summation methods and
understanding how the accuracy of the used approximation depends on the number of terms
present in the series. According to \cite{Benber}, the oscillator ground state energy
$E_{0}(g)$ can be represented as a perturbation series in the
anharmonicity constant $g$ whose first few terms have the form
\begin{equation}
E_{0}(g)=\sum\limits_{n=0}^{\infty}c_{n}g^{n}=1+\frac{3}{4}g-\frac{21}{16}g^{2}+\frac{333}{64}g^{3}-\frac{30885}{1024}g^{4}+\frac{916731}{4096}g^{5}-\frac{65518401}{32768}g^{6}+\dots,
\label{f_E_0}
\end{equation}
and exact values of $E_{0}$ depending on $g$ are given in \cite{Cizek}.
It was shown in \cite{Benber} that the general term of series (\ref{f_E_0}) increases
factorially as in expression (\ref{f_beta}).

The PB method is based on the idea that a series of form (\ref{f_E_0}) can be written as
the integral
\begin{equation}
f(g)=\int\limits_{0}^{\infty}e^{-t}B(gt)dt,\qquad B(g)=\sum\limits_{n=0}^{\infty}B_{n}g^{n},\qquad B_{n}=\frac{c_{n}}{n!}.
\label{f_f}
\end{equation}

In this expression, $B(g)$ is called the Borel image.
According to asymptotic expansion (\ref{f_beta}), the Borel image
converges in a disk of radius $1/a$.
We then apply the Pad\'{e} approximation to $B(g)$, which consists in using
a rational function of the form
\begin{equation}
[L/M]=\frac{\sum_{i=0}^{L}a_{i}g^{i}}{\sum_{j=0}^{M}b_{j}g^{j}}\qquad (M\geq 1),
\label{f_[L/M]}
\end{equation}
whose expansion in a Taylor series (in a neighborhood of the point $g=0$) coincides with
the expansion of the Borel image as closely as possible.
A function of form (\ref{f_[L/M]}) has $L+1$ coefficients in the numerator and
$M+1$ coefficients in the denominator. The entire set of coefficients is thus determined
up to the common factor; to be definite, we usually set $b_{0}=1$.
As a result, we obtain the total set of $L+M+1$ free parameters.
This means that the coefficients of the expansion of the function $[L/M]$ in a Taylor series must generally
coincide with the corresponding coefficients of series (\ref{f_E_0}).
If the number of terms N in the series is finite,
then the approximants $[L/M]$ must satisfy the condition $L+M\leq N$.

A generalization of the PB method is the PBL method for which formulas (\ref{f_f})
can be written in the extended form as
\begin{equation}
f(g)=\int\limits_{0}^{\infty}t^{b}e^{-t}B(gt)dt,\qquad B(g)=\sum\limits_{n=0}^{\infty}B_{n}g^{n},\qquad B_{n}=\frac{c_{n}}{\Gamma(n+b+1)},
\label{f_f_b}
\end{equation}
where $b$ is an arbitrary parameter. In the simplest realization of the method, we consider
the values $b=0$ and $b=1$, for example,
and then analyze the variations in the series approximations resulting
from variations in the parameter $b$, which permits estimating the accuracy of approximations
obtained by the PB and PBL methods. In our case where the PBL method is applied to the series
determining the characteristics of the critical behavior of homogeneous and disordered systems,
for a test series, we choose series (\ref{f_E_0}) corresponding to the exactly solvable problem
of calculating the anharmonic oscillator energy
with the asymptotic convergence of the series, which is similar to that of
series (\ref{f_gamma_d=2}) and (\ref{f_gamma_d=3}) in the theory
of critical phenomena. Therefore, applying the PBL method to series (\ref{f_gamma_d=2}) and (\ref{f_gamma_d=3})
in the theory of critical phenomena, we plan to use the values of the parameter $b$ that
lead to the best approximations of series (\ref{f_E_0}).

\begin{figure}[t]
\centering
\includegraphics[width=\textwidth]{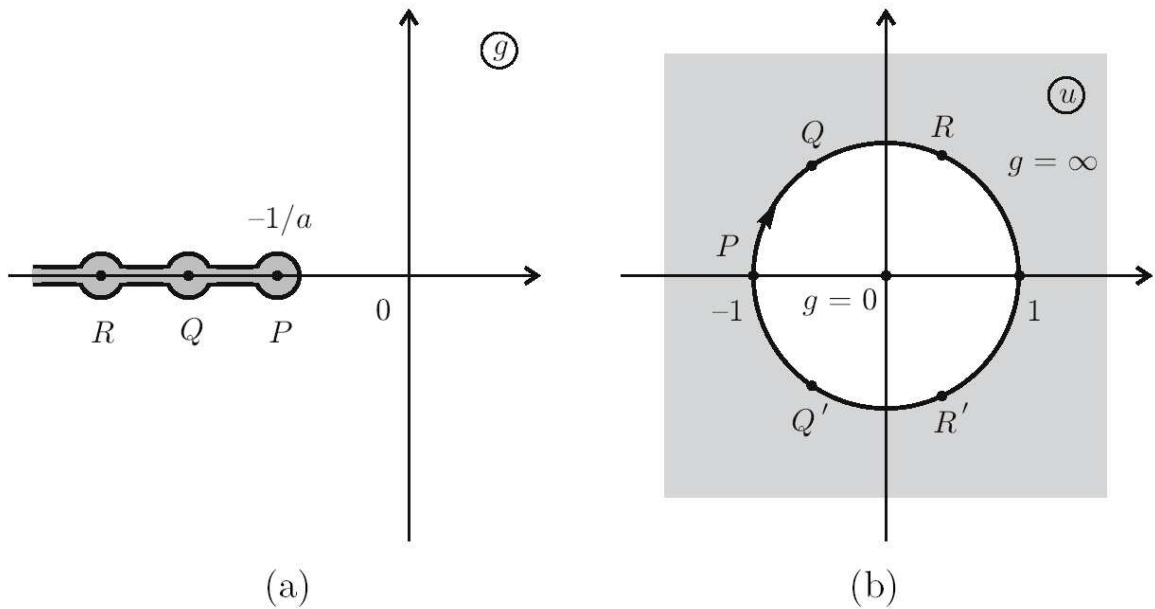}
\caption{\label{fig:1_conf} Domain of the Borel image analyticity on the complex plane with the cut
$(-\infty,-1/a)$ along the real axis \textbf{(a)} and its conformal mapping to the
unit disk \textbf{(b)}.}
\end{figure}

\begin{figure}[t]
\centering
\includegraphics[width=0.75\textwidth]{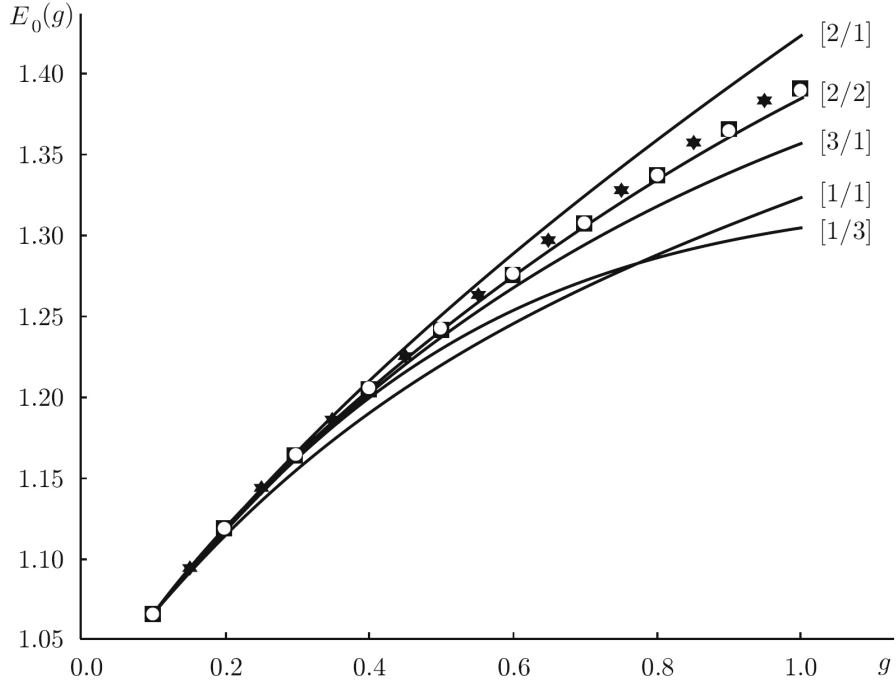}
\caption{\label{fig:2_conf} Comparison of the exact values of the anharmonic oscillator ground state energy $E_{0}$ (black
squares) for different values of the anharmonicity constant g with the results obtained by the PB
summation method for different $[L/M]$ (solid lines), the PBL method with the parameter $b=2.221426$
for the approximant $[1/1]$ (white circles), and the CPB method with the constant $a=3$ (stars).}
\end{figure}

We consider the main principles of applying the CM method to the Borel image (for $b=0$).
According to asymptotic expansion (\ref{f_beta}), the Borel image is an analytic function
in the complex plane of $g$ with the
cut from $-1/a$ to $-\infty$ (Fig.~\ref{fig:1_conf}a). Because the integration in formulas (\ref{f_f_b}) is
performed along the entire real axis, we must analytically continue $B(g)$ to arbitrary
complex $g$ beyond the convergence disk $\left|g\right|<1/a$,
and this continuation is realized by the conformal map $g=w(u)$ with
\begin{equation}
w(u)=\frac{4}{a}\frac{u}{(1-u)^{2}},\qquad u=\frac{(1+ag)^{1/2}-1}{(1+ag)^{1/2}+1},
\label{f_w}
\end{equation}
which takes the plane with a cut to the unit disk $\left|u\right|<1$ (Fig.~\ref{fig:1_conf}b).
The representation of $B(g)$ as a series in $u$ thus leads to the convergent series
for any g:
\begin{equation}
B(g)=\sum\limits_{n=0}^{\infty}\left.B_{n}g^{n}\right|_{g=w(u)}\qquad \Longrightarrow\qquad B(u)=\sum\limits_{n=0}^{\infty}U_{n}u^{n}.
\label{f_B}
\end{equation}
Indeed, all possible singular points $P,Q,R,\dots$ of the function $B(g)$ lie along the cut,
and their images $P,Q,Q',R,R',\dots$ lie on the boundary of the disk $\left|u\right|=1$.
Hence, the second of series (\ref{f_B}) converges for all $\left|u\right|<1$.
The relationship between the coefficients $U_{n}$ and $B_{n}$ can be expressed as
\begin{equation}
U_{0}=B_{0},\qquad U_{n}=\sum\limits_{m=1}^{n}B_{m}\left(\frac{4}{a}\right)^{m}C_{n+m-1}^{n-m}\qquad (n\geq 1).
\label{f_U_n}
\end{equation}
These formulas solve the above problem as follows: the Borel image $B(u(g))$
converges for any singular points $g=\infty, g=-1/a$ and $g=g_{0}$ with $g_{0}\in(-\infty,-1/a)$,
and its coefficients $U_{n}$ are related are related to the
original coefficients $c_{n}$ and the parameter $b$ by linear transformation (\ref{f_U_n}) (see formulas (\ref{f_f_b})).
Because we have $|u|<1$ for any $g$, the Pad\'{e} approximation can be effectively applied to the series in the variable $u$.
This procedure is called the conformal Pad\'{e}-Borel (CPB) method.

\begin{table}[t]
\footnotesize
  \caption{ Sums of squared deviations of the values of the oscillator energy $E_{0}$ calculated using the PB, PBL,
and CPB methods from the exact values of $E_{0}$ (the gaps show that the Borel image has a pole
when the corresponding approximant is used). }
  \begin{center}
    \begin{tabular}{|c|cccccccc|}
      \hline
          [L/M]   &  [1/1] &  [1/2] &  [2/1] &  [1/3] &      [2/2]     &  [3/1] &  [1/4] &  [2/3] \\
      \hline
      $S_\text{PB}^{2}$&0.013923&    -   &0.002703&0.016687&    0.000105    &0.002535&    -   &    -   \\
      \hline
      $     b    $&2.221426&1.582184&1.466092&1.276639&    1.131538    &1.194466&2.016677&1.050884\\
      $          $&        &3.441754&        &3.879020&                &        &4.083434&        \\
      $S_\text{PBL}^{2}$&0.000011&0.000006&0.000003&0.000007&$3\cdot 10^{-7}$&0.000001&0.000010&0.000044\\
      $          $&        &0.000010&        &0.000008&                &        &0.000007&        \\
      \hline
      $S_\text{CPB}^{2}$&0.000175&0.041795&0.000836&0.000307&    0.000192    &    -   &    -   &0.000011\\
      \hline
      \hline
          [L/M]   &      [3/2]     &  [4/1] &   [0/6]  &  [1/5] &      [2/4]     &      [3/3]     &      [4/2]     &  [5/1] \\
      \hline
      $S_\text{PB}^{2}$&    0.000001    &0.004142& 2.609852 &0.031233&        -       &        -       &    0.000247    &0.000833\\
      \hline
      $     b    $&    1.005754    &1.068052&$\nexists$&2.445345&    0.970348    &    0.959599    &    0.964316    &1.006032\\
      $          $&                &        &          &4.194370&                &    2.003840    &                &        \\
      $S_\text{PBL}^{2}$&$3\cdot 10^{-7}$&0.000004&          &0.000007&$2\cdot 10^{-7}$&$2\cdot 10^{-7}$&$2\cdot 10^{-7}$&0.000008\\
      $          $&                &        &          &0.000006&                &$2\cdot 10^{-7}$&                &        \\
      \hline
      $S_\text{CPB}^{2}$&        -       &    -   & 0.000064 &    -   &$3\cdot 10^{-7}$&$5\cdot 10^{-7}$&    0.003932    &0.000003\\
      \hline
    \end{tabular}
  \end{center}
  \label{t_S2}
\end{table}

\begin{table}
\footnotesize
  \caption{Values of the fixed point $g^*$ and the dynamical critical exponent $z$ for the two- and
three-dimensional Ising models obtained by the PB, PBL, and CPB methods.}
  \begin{center}
    \begin{tabular}{|c|cc|ccccccc|}
      \hline
         &  & [L/M] & [1/1]& [1/2]& [2/1]& [1/3]& [3/1]& [1/4]& [2/3]\\
      \hline
         &PB&$g^{*}$&2.1971&   -  &1.8210&1.8271&1.7702&      &      \\
         &  &$  z  $&2.1262&      &2.0868&2.0873&2.0820&      &      \\
      \hspace*{1mm}d=2\hspace*{1mm}&PBL&$g^{*}$&1.6978&1.7098&   -  &1.7174&1.6849&      &      \\
         &  &$  z  $&2.0753&2.0764&      &2.0771&2.0742&      &      \\
         &CPB&$g^{*}$&2.1448&   -  &1.9137&   -  &1.8193&      &      \\
         &  &$  z  $&2.1270&      &2.1018&      &2.0922&      &      \\
      \hline
         &PB&$g^{*}$&1.5508&   -  &1.4314&1.4230&1.4175&1.4427&1.4753\\
         &  &$  z  $&2.0202&      &2.0172&2.0170&2.0169&2.0175&2.0183\\
      d=3&PBL&$g^{*}$&1.3972&1.4019&1.4671&1.4056&1.4049&1.4073&   -  \\
         &  &$  z  $&2.0166&2.0165&      &2.0166&2.0166&2.0166&      \\
         &CPB&$g^{*}$&1.5385&   -  &1.4552&   -  &1.4267&1.4253&   -  \\
         &  &$  z  $&2.0452&      &2.0393&      &2.0374&2.0373&      \\
      \hline
      \hline
         &  & [L/M] & [3/2]& [4/1]& [1/5]& [2/4]& [3/3]& [4/2]& [5/1]\\
      \hline
         &PB&$g^{*}$&1.4234&1.4292&   -  &1.4263&   -  &1.4200&1.3854\\
         &  &$  z  $&2.0170&2.0172&      &2.0171&      &2.0169&2.0161\\
      d=3&PBL&$g^{*}$&1.4234&1.4355&1.4082&   -  &1.4221&1.4208&1.3848\\
         &  &$  z  $&2.0170&2.0173&2.0167&      &2.0170&2.0170&2.0161\\
         &CPB&$g^{*}$&1.4242&1.4245&   -  &   -  &   -  &   -  &1.4147\\
         &  &$  z  $&2.0372&2.0373&      &      &      &      &2.0366\\
      \hline
    \end{tabular}
  \end{center}
  \label{t_z_d}
\end{table}

In Fig.~\ref{fig:2_conf}, we compare the approximations applied to the series for the anharmonic oscillator ground
state energy with the exact values of $E_{0}$ obtained using the PB method with different types of the Pad\'{e}
approximant $[L/M]$ and the PBL and CPB methods (for $a=3$ \cite{Benber}) with the approximant $[1/1]$. Fig.~\ref{fig:2_conf}
shows that the approximation error of the PB method increases strongly as $g$ increases, although the
situation becomes better as the number $N$ of the series terms taken into account increases and the diagonal
and nearly diagonal approximants are used. But the PBL method for $b = 2.221426$ (in expression (\ref{f_f_beta})) and
the CPB method even for the approximant $[1/1]$ give results close to the exact values of $E_0$ in the considered
interval of the variable $g$. These results can be compared in accuracy only with the results obtained using
the "best" diagonal approximant $[2/2]$ in the PB method.

In realizing the PBL method, it is more convenient to perform the integral Borel transformation as
\begin{equation}
f(g)=\int\limits_{0}^{\infty}dte^{-t}B(gt^{b}),\qquad B(g)=\sum\limits_{n=0}^{\infty}B_{n}g^{n},\qquad B_{n}=\frac{c_{n}}{\Gamma(bn+1)},
\label{f_f_beta}
\end{equation}
where the parameter $b$ for all $[L/M]$ is determined by the requirement that the mean squared deviation of
approximations from the exact values of $E_{0}$ be minimum on the entire interval of~$g$. Using expression (\ref{f_f_beta})
allows approximating the exact value of $E_0$ much better than using (\ref{f_f_b}). Moreover, this transformation, in
contrast to transformation (\ref{f_f_b}), allows avoiding the appearance of singularities in the integrand when the
abovementioned variational method for choosing the values of $b$ is used.

Table~\ref{t_S2} presents values of the sum of squared deviations $S^{2}$ for the values of the oscillator energy $E_{0}$
when different approximations of the exact values of $E_{0}$ are used in the PB, PBL, and CPB methods. In
Table~\ref{t_S2}, we do not present the results obtained using approximants of the form $[0/N]$, because it was found
that for even $N$, using them leads to very large deviations of the series sum from the exact solution and the
deviation $S^{2}$ does not have a minimum in $b$ (only a small decrease in $S^{2}$ is observed as $b$ increases), while
for odd $N$, singularities unremovable for any value of $b$ appear in the Borel image corresponding to this
approximant. For each of the approximants of the form $[1/N - 1]$, we found two values of the parameter
$b$ characterized by close values of the deviation $S^{2}$. Comparing the given values of $S^{2}$ clearly shows that
the accuracy of the PBL method is much higher than that of the PB method. The PB method permits
obtaining results comparable in accuracy to the results obtained by the PBL method only starting from the
fifth-order nearly diagonal approximants. This makes the PBL method preferable for analyzing the short
series obtained in the description of the critical dynamics of homogeneous systems (\ref{f_gamma_d=2}), (\ref{f_gamma_d=3}) and disordered
systems (\ref{f_gamma}). We also note that the CPB method is second in accuracy among the methods considered
and is as good in accuracy as the PBL method starting from the fifth-order approximants in $N$ (except for
the case of the approximant $[4/2]$). The PB method, although it is below the CPB method in accuracy for
short series, permits obtaining comparable results only for the diagonal approximant $[2/2]$ and the nearly
diagonal approximants $[3/2]$ and $[4/2]$.

We now consider how the above methods can be used to calculate the dynamical critical exponent $z$
for the homogeneous Ising model describing the critical behavior of systems with the dimensions $d=2$ and
$d=3$. As previously noted, the exponent $z$ is determined when the functions $\beta(g)$ and $\gamma_{\lambda}(g)$ are used. The
fixed point $g^{*}$ of the RG transformations is found from the condition $\beta(g^{*})=0$. In Table~\ref{t_z_d}, we present the
values of the fixed point $g^{*}$ and the dynamical critical exponent $z$ for the two- and three-dimensional Ising
models, which were obtained by the PB, PBL (for the values of the parameter $b$ related to the corresponding
approximants in the oscillator problem), and CPB methods, where, according to \cite{Baker,LeGuillou}, the constant $a$ for the
last method was taken to be $0.238659217$ for $d=2$ and $0.14777422$ for $d=3$. In the PB and PBLmethods,
the series for the function $2+\gamma_{\lambda}$ was summed using the fourth-order approximant $[3/1]$, which is the best
(summable) approximant for these methods, while in the CPB method, all the approximants other than
$[N/0]$ turned out to be nonsummable, and only the simple CM method was therefore realized.

The averaging procedure for the results obtained using the PB method with the approximants of order
$N\geq 4$ taken into account allows obtaining the following values of the charge $g^{*}$ at the fixed point and of
the exponent $z$:
\begin{eqnarray}
g^{*}=1.7987\pm 0.0201,\qquad z=2.0847\pm 0.0019\ (d=2);\nonumber\\
g^{*}=1.4270\pm 0.0074,\qquad z=2.0171\pm 0.0002\ (d=3),\nonumber
\end{eqnarray}
Using the PBL method gives
\begin{eqnarray}
g^{*}=1.7012\pm 0.0115,\qquad z=2.0757\pm 0.0010\ (d=2);\nonumber\\
g^{*}=1.4125\pm 0.0046,\qquad z=2.0168\pm 0.0001\ (d=3)\ \nonumber
\end{eqnarray}
Using the CPB method gives
\begin{eqnarray}
g^{*}=1.8193,\phantom{\pm 0.0000\ }\qquad z=2.0922\phantom{\pm 0.0000\,}\ (d=2); \nonumber\\
g^{*}=1.4231\pm 0.0019,\qquad z=2.0372\pm 0.0001\ (d=3). \nonumber
\end{eqnarray}

We note that all the obtained values of the charge $g^{*}$ at the fixed point for $d=3$ agree well with the result
 $g^{*}=1.416\pm 0.005$ obtained by Le Guillou and Zinn-Justin \cite{LeGuillou}.

\begin{table}
\footnotesize
  \caption{Results obtained by the PB, PBL, and CPB methods in calculating the fixed points and the
dynamical critical exponent $z$ for the disordered Ising model}
  \begin{center}
    \begin{tabular}{|cc||c|c|cc|ccc|cc|}
      \hline
        &       &  N=2  &  N=3  &  N=4  &       &  N=5  &       &       &  N=6  &       \\
        & [L/M] & [1/1] & [1/2] & [2/2] & [3/1] & [1/4] & [3/2] & [4/1] & [4/2] & [5/1] \\
      \hline
      PB&$v^{*}$&-0.5868&-0.6894&-0.7383&-0.6991&-0.6394&-0.7146&-0.7201&-0.7125&       \\
        &$g^{*}$& 2.3900& 2.2402& 2.2743& 2.2411& 2.2497& 2.2640& 2.2705& 2.2583&       \\
        &$  z  $& 2.1509& 2.1756& 2.1878& 2.1780& 2.1633& 2.1819& 2.1833& 2.1814&       \\
      \hline
      PBL&$v^{*}$&-0.6862&-0.7001&-0.7000&-0.7142&-0.6900&-0.7142&-0.7089&-0.7141&       \\
        &$g^{*}$& 2.3895& 2.2499& 2.2499& 2.2459& 2.2601& 2.2638& 2.2646& 2.2600&       \\
        &$  z  $& 2.1757& 2.1787& 2.1787& 2.1822& 2.1762& 2.1823& 2.1809& 2.1822&       \\
      \hline
      CPB&$v^{*}$&       &-0.6800&-0.6700&-0.6808&-0.7302&-0.7239&       &-0.7321&-0.7321\\
        &$g^{*}$&       & 2.1654& 2.1342& 2.2562& 2.3079& 2.2671&       & 2.2763& 2.2799\\
        &$  z  $&       & 2.1716& 2.1691& 2.1710& 2.1844& 2.1831&       & 2.1854& 2.1853\\
      \hline
  \end{tabular}
  \end{center}
  \label{t_z}
\end{table}

Averaging the above values obtained by the different methods finally gives
\begin{eqnarray}
g^{*}=1.4209\pm 0.0035,\qquad z=2.0237\pm 0.0055\ (d=3); \nonumber\\
g^{*}=1.7731\pm 0.0297,\qquad z=2.0842\pm 0.0039\ (d=2).\nonumber
\end{eqnarray}
In the case where the critical behavior of disordered systems is described, the theory series for the
functions $\beta_{g}$, $\beta_{v}$ and $\gamma_{\lambda}$ (see (\ref{f_gamma})) are two-parameter series:
\begin{equation}
F(v,g)=\sum\limits_{i,j}C_{i,j}v^{i}g^{j}\qquad (i,j=0,1,2,\dots).
\label{f_F}
\end{equation}
The methods described above cannot be applied to such series directly. They must be modified for the case
of several variables. One that can be thus modified is called the $\lambda$-method. The essence of this method
consists in introducing a generalized series
\begin{equation}
\widetilde{F}(v,g;\lambda)=\sum\limits_{n=0}^{\infty}\widetilde{C}_{n}(v,g)\lambda^{n},\qquad \widetilde{C}_{n}(v,g)=\sum\limits_{i,j}C_{i,j}v^{i}g^{j}\delta_{i+j,n},
\label{f_F_lambda}
\end{equation}
which converges to the original series (\ref{f_F}) for $\lambda=1$. We can assume that such a series depends on the
single variable $\lambda$ and its coefficients are functions of $u$ and $v$. The Borel transform of series (\ref{f_F}) has the
form
In the $\lambda$-method, the Borel image is reduced to a function depending only on the single variable $\lambda$, the Pad\'{e}
approximant method is then applied to this function, and the integral is calculated.
In the CM method applied to two-parameter series, we use the change  $\delta=v/u$, which permits rewriting
series (\ref{f_F}) as
\begin{equation}
\left.F(v,g)\right|_{g=\delta v}=\sum\limits_{n=0}^{\infty}\widetilde{C}_{n}(\delta)v^{n}.
\label{f_F_delta}
\end{equation}
The Borel transform then becomes
\begin{equation}
F(v,\delta)=\int\limits_{0}^{\infty}dte^{-t}B(vt^{b},\delta),\qquad B(v,\delta)=\sum\limits_{n=0}^{\infty}\frac{\widetilde{C}_{n}(\delta)}{\Gamma(nb+1)}v^{n}.
\label{f_F_delta_beta}
\end{equation}
The conformal map is determined by the formula
\begin{equation}
u(v)=\frac{\sqrt{1+\alpha(\delta)v}-1}{\sqrt{1+\alpha(\delta)v}+1}.
\label{f_u}
\end{equation}
When the CM method is applied to the series of the critical behavior theory for disordered systems, the
problem becomes more complicated because the asymptotic behavior of the series is unknown in this case.
In \cite{Pelissetto}, for the Ising model, it was proposed to take $\alpha(\delta)$ in the form
\begin{eqnarray}
\alpha(\delta)=a(9/8+\delta)\qquad(\delta<0,\delta>4),\nonumber
\end{eqnarray}
where $a=0.14777422$ for the three-dimensional Ising model.

We now consider the application of these two methods to series (\ref{f_gamma}), which characterize the nonequilibrium
critical behavior of the disordered Ising model. In Table~\ref{t_z}, we present the results obtained by these
summation methods in calculating the coordinates of the fixed point, which determines the critical behavior
of disordered systems, and the exponent $z$. In the PB and PBL methods, we summed the series for the
function $2+\gamma_{\lambda}$ using the third-order approximant $[2/1]$, which is the best for these methods, and in the
CPB method, we used the approximant $[1/2]$. It follows from Table~\ref{t_z} that all these methods give close
values for the vertices at the fixed points and for the exponent $z$.

We now apply the averaging procedure to the obtained results using all the values of the approximants
$[L/M]$ except $[1/1]$. The averages of the fixed-point coordinates and the exponent z obtained using the PB
method are
$$
v^{*}=-0.7019\pm 0.0111,\qquad g^{*}=2.2569\pm 0.0048,\qquad z=2.1788\pm 0.0027,
$$
Using the PBL method gives
$$
v^{*}=-0.7059\pm 0.0033,\qquad g^{*}=2.2563\pm 0.0026,\qquad z=2.1802\pm 0.0008\,
$$
Using the CPB method gives
$$
v^{*}=-0.7070\pm 0.0100,\qquad g^{*}=2.2410\pm 0.0227,\qquad z=2.1786\pm 0.0026.
$$
The close values of the fixed-point coordinates and the exponent $z$ obtained by different methods and
practically coinciding within the accuracy confirm the reliability of the obtained results. Averaging these
values obtained by different methods finally gives
$$
v^{*}=-0.7049\pm 0.0013,\qquad g^{*}=2.2514\pm 0.0042,\qquad z=2.1792\pm 0.0004.
$$
The dynamical critical exponent $z$ was calculated in \cite{Prudnikov_1998} using the Chisholm--Borel method and $z=2.1653$
was obtained. This value of the exponent agrees rather well with the results obtained here.

\section{Conclusion}
In this paper, we first successively applied the PB, PBL, and CM summation methods with subsequent
use of the Pad\'{e} approximation (the CPB method) to determine the values of the dynamical critical exponent
$z$ for homogeneous two- and three-dimensional systems described by the Ising model and for disordered
Ising-like systems. Our analysis was based only on the now available short series of the theory corresponding
to the results of the four-loop description of the critical dynamics of homogeneous systems and the three-loop
description of disordered systems \cite{Prudnikov_1997,Prudnikov_1998}.

We compare the obtained results with the experimental data and the results of Monte Carlo computer
simulations of the critical dynamics. In \cite{Belanger}, the critical dynamics of the homogeneous Ising antiferromagnet
$FeF_{2}$ was investigated experimentally, and the value $z=2.1\pm 0.1$ was obtained, which in the framework of a
sufficiently wide interval of measurement error for $z$ does not contradict our results of the field theory calculations
using the summation methods. The Monte Carlo numerical investigations of the critical dynamics of
the homogeneous three-dimensional Ising model for different data given in the literature gave the following
values of the dynamical critical exponent: $z=1.99\pm 0.02$ \cite{Pearson}, $z=1.97\pm 0.08$ \cite{Prudnikov_1993},
$z=2.04\pm 0.03$ \cite{Wansleben}, $z=2.04\pm 0.01$ \cite{Gropengiessen} and
$z=2.032\pm 0.004$ \cite{Grassberger}.
These values of the exponent $z$ agree well with the results
obtained here and confirm that the our predicted value $z=2.0237\pm 0.0055$ is sufficiently reliable.

As for the two-dimensional homogeneous Ising model, the numerical investigations of its critical
dynamics, in contrast to similar investigations of the three-dimensional model, are characterized by a
much wider range of the values obtained for the dynamical critical exponent: $z=2.14\pm 0.02$ \cite{Kalle}, $z=2.13\pm 0.03$ \cite{Williams},
$z=2.076\pm 0.005$ \cite{Mori}, $z=2.24\pm 0.04$ \cite{Poole}, $z=2.24\pm 0.07$ \cite{Prudnikov_95},
$z=2.16\pm 0.04$ \cite{Wang} and $z=2.1667\pm 0.0005$ \cite{Nightingale}.
This range of $z$ is wide mainly because the critical behavior of the two-dimensional
Ising model is characterized by both much greater amplitudes of the order parameter fluctuations in the
critical domain (compared with the three-dimensional model) and more significant effects of the critical delay.
The high-temperature expansion method used to describe the critical dynamics of the two-dimensional
Ising model gives the value $z=2.125$ \cite{Racz}. The values of the exponent $z$ obtained for the two-dimensional
Ising model in the cited papers are in the rather wide interval $2.07\leq z \leq 2.31$, while the values obtained
here are on the lower boundary of this interval. This makes the results obtained here more valuable as
reference points in future studies.

We now compare the values of the dynamical critical exponent for disordered systems calculated in this
paper with the results of computer simulations of the critical dynamics of the three-dimensional disordered
Ising model: $z=2.19\pm 0.07$ for systems with spin concentration $p = 0.95$,
$z=2.20\pm 0.08$ for $p=0.8$, $z=2.58\pm 0.09$ for $p=0.6$
and $z=2.65\pm 0.12$ for $p=0.4$ \cite{Prudnikov_1993};
$z=2.16\pm 0.01$ for $p=0.95$, $z=2.232\pm 0.004$ for $p=0.9$,
$z=2.38\pm 0.01$ for $p=0.8$ and $z=2.93\pm 0.03$ for $p=0.6$ \cite{Heuer}.
These results
of critical dynamics simulations agree rather well with the results of field theory calculations using the
summation methods only for weakly disordered systems with $p\geq 0.8$, while a noticeable difference between
the results is observed for strongly disordered systems. We note that the results of the RG description of
the critical behavior of disordered systems in this case hold only in the domain of weak disorder. To explain
the dependence of the exponent $z$ on the value of the structural disorder observed in computer simulations,
the hypothesis of graduated universality was proposed in \cite{Prudnikov_1993}, according to which several types of different
critical behavior can be observed in systems with spin concentrations greater than the spin percolation
threshold depending on whether there exists only one spin percolation cluster in the system as in the case
of weakly disordered systems or there is an admixture percolation cluster in addition to the spin percolation
cluster as in the case of strongly disordered systems with transient regimes between the domains.

Starting from the concept that the critical behavior of disordered systems is universal and that the
asymptotic value (as $L \to \infty$) of the exponent z is independent of the degree of disorder, the author of \cite{Heuer}
obtained the asymptotic value $z=2.4\pm 0.1$ using the effective values of the exponent listed above. But
this value of the exponent is strongly inconsistent with the results obtained here. A numerical study of the
critical dynamics of the three-dimensional Ising model with the spin concentration varying in a wide interval
was analyzed in \cite{Parisi}. Assuming that the critical behavior of disordered systems is universal, the authors
of \cite{Parisi} obtained the asymptotic value of the exponent $z=2.62\pm 0.07$ taking the effects of influence of the
leading corrections to the scaling dependence for the dynamical susceptibility of the system into account.
In this case, the value of the exponent of the scaling correction, $\omega = 0.50\pm 0.13$, obtained in \cite{Parisi} is strongly
inconsistent with the results of the field theory calculations of the static critical exponents, which were done
using the summation methods in \cite{Pelissetto} and gave $\omega = 0.25\pm 0.10$. This value is also inconsistent with the
results of numerical investigations of the static critical behavior of the same model \cite{Ballesteros} also done taking
the effects of influence of the leading corrections to the scaling dependence of thermodynamic quantities
and the correlation functions with $\omega = 0.37\pm 0.06$ into account. In the approximations realized in \cite{Parisi}, the
results for weakly disordered systems were characterized by the largest errors. We also note that the value
of the dynamical critical exponent z obtained in \cite{Parisi} is even more strongly inconsistent with our results
here.

The critical dynamics of the disordered Ising antiferromagnet $Fe_{0.46}Zn_{0.54}F_{2}$ was studied experimentally
in \cite{Belanger}, where $z=1.7\pm 0.2$ was obtained. This result conflicts with both the results of the field
theory calculations using the summation methods and the above results of computer simulations. This
inconsistency can possibly be explained by a high concentration of nonmagnetic admixture atoms in the
sample under study and by the presence of large-scale inhomogeneities in this sample, which significantly
affects the characteristics of the nonequilibrium critical behavior. But in the other experimental paper
concerning the critical dynamics of weakly diluted Ising magnet $Fe_{0.9}Zn_{0.1}F_{2}$ \cite{Rosov}, the value of the dynamical
critical exponent $z=2.18\pm 0.10$ was obtained as the result of high-precession measurement of the
dynamical widening of M$\ddot{o}$ssbauer lines using the M$\ddot{o}$ssbauer spectroscopy method. This value agrees very
well with the results of our calculations. Nevertheless, there is a strong need for additional investigations
of the critical dynamics of disordered Ising-like systems, both experimentally and numerically. The results
of calculations of the dynamical critical exponent for disordered systems obtained here will be reference
points in such investigations.

\begin{acknowledgments}
This work was supported by the Russian Foundation for Basic Research through Grants (No.~04-02-17524, No.~04-02-39000 and No.~05-02-16188),
and by Grant No. MK-8738.2006.2 of Russian Federation President.
\end{acknowledgments}

\end{document}